# Effects of Mismatch Strain and Substrate Surface Corrugation on Morphology of Supported Monolayer Graphene

Zachary H. Aitken and Rui Huang*

Department of Aerospace Engineering and Engineering Mechanics, The University of Texas at Austin, Austin, TX 78712

**Abstract**

Graphene monolayers supported on oxide substrates have been demonstrated with superior charge mobility and thermal transport for potential device applications. Morphological corrugation can strongly influence the transport properties of the supported graphene. In this paper, we theoretically analyze the morphological stability of a graphene monolayer on an oxide substrate, subject to van der Waals interactions and in-plane mismatch strains. First, we define the equilibrium separation and the interfacial adhesion energy as the two key parameters that characterize the van der Waals interaction between a flat monolayer and a flat substrate surface. By a perturbation analysis, a critical compressive mismatch strain is predicted, beyond which the graphene monolayer undergoes strain-induced instability, forming corrugations with increasing amplitude and decreasing wavelength on a perfectly flat surface. When the substrate surface is not perfectly flat, the morphology of graphene depends on both the amplitude and the wavelength of the surface corrugation. A transition from conformal (corrugated) to non-conformal (flat) morphology is predicted. The effects of substrate surface corrugation on the equilibrium mean thickness of the supported graphene and the interfacial adhesion energy are analyzed. Furthermore, by considering both the substrate surface corrugation and the mismatch strain, it is found that, while a tensile mismatch strain reduces the corrugation amplitude of graphene, a corrugated substrate surface promotes strain-induced instability under a compressive strain. These theoretical results suggest possible means to control the morphology of graphene monolayer on oxide substrates by surface patterning and strain engineering.

*Corresponding author. Email: ruihuang@mail.utexas.edu.

## I. Introduction

The stability of two-dimensional (2-D) lattice structure of suspended graphene sheets has been attributed to the intrinsic microscopic corrugation in the third dimension [1]. Numerical simulations have suggested different mechanisms for the observed corrugations in graphene, including thermal fluctuation [2] and molecular absorption [3]. Supported on a substrate, corrugation of monolayer graphene has been observed to partly conform to the substrate surface [4-6]. The interfacial interaction between graphene and its substrate, which varies from strong chemical bonds for epitaxial graphene on a single-crystal substrate [7-11] to weak van der Waals forces for mechanically exfoliated graphene on an amorphous substrate (e.g., silicon dioxide or $SiO_2$) [4-6], plays a critical role in determining the morphology of supported graphene. So does the intrinsic elastic stiffness of the monolayer graphene [12, 13], for both in-plane and bending deformation. Furthermore, a mismatch strain between graphene and the supporting substrate could also affect the morphology of both fully and partly supported graphene [14]. Such a mismatch strain could result either from a differential thermal expansion between graphene and an oxide substrate [14] or from the lattice mismatch between an epitaxial graphene and its crystalline substrates [8, 11]. In general, it is important to understand the quantitative nature of both the intrinsic and extrinsic corrugation in graphene because the morphology can strongly influence the physical properties of graphene, such as the electronic structures [15, 16] and thermal conductivity [17, 18].

In the present study, we theoretically analyze the effects of substrate surface corrugation and mismatch strain on the morphology of a graphene monolayer supported on an oxide substrate. The effect of surface corrugation has been considered in a recent study [19], where the van der Waals interaction energy between a monolayer graphene and its substrate was numerically calculated using a Monte Carlo method. Here we present an analytical approach that explicitly relates the interaction energy to the surface corrugation and the interfacial properties. Moreover, the effect of mismatch strain is considered, which predicts a strain-induced instability under a compressive strain and reduced corrugation under a tensile strain. Together, these theoretical results suggest that the morphology of graphene can be tunable via both substrate surface patterning and strain engineering.

**II. Graphene on a flat surface**

In this section, we consider van der Waals interactions and strain-induced instability of monolayer graphene on an oxide substrate with a perfectly flat surface. The effects of surface corrugation are analyzed in Section III.

A. *van der Waals Interaction*

The free energy of van der Waals interaction between a monolayer and a substrate can be obtained from the pairwise potential energy for the atom-atom interaction based on the Hamaker summation method [20]. In the present study, we take the standard form of the Lennard-Jones potential for the pairwise interaction between a carbon atom in graphene and a substrate atom, namely,

$$W_{LJ}(r) = -\frac{C_1}{r^6} + \frac{C_2}{r^{12}}, \qquad (1)$$

where $r$ is the distance between the two atoms, $C_1$ and $C_2$ are the constants for the attractive and repulsive interactions, respectively. Assuming an effectively homogeneous substrate, we sum (integrate) the energy between one carbon atom and all the atoms in the substrate to obtain an atom-surface potential for each carbon atom near the surface. Next, summing up the atom-surface potential for all the carbon atoms in the monolayer graphene (flat or corrugated) results in the monolayer-surface interaction energy, namely

$$W = \int_{A_g} \int_{V_s} W_{LJ} \rho_s \rho_g dV_s dA_g, \qquad (2)$$

where $\rho_s$ is the number of atoms per unit volume of the substrate, $\rho_g$ is the number of carbon atoms per unit area of the graphene monolayer, $V_s$ is the substrate volume, and $A_g$ is the area of the graphene monolayer. By the integration in Eq. (2), the substrate has been treated as a three-dimensional continuum and the graphene monolayer a two-dimensional continuum membrane. Such a treatment is sufficient for the present study where the characteristic length scale (e.g., corrugation wavelength) is relatively large compared to the interatomic bond lengths in both the graphene and the substrate.

Near a flat substrate surface, the atom-surface interaction potential has been well documented [20], based on which the interaction potential between a flat monolayer and a flat substrate surface can be written in an analytic form

$$U_{vdW}(z) = -\Gamma_0 \left[ \frac{3}{2} \left( \frac{h_0}{z} \right)^3 - \frac{1}{2} \left( \frac{h_0}{z} \right)^9 \right], \qquad (3)$$

where $U_{vdW}$ is the monolayer-surface interaction energy per unit area, $z$ is the distance between the monolayer and the substrate surface (Fig. 1a), $h_0$ is the equilibrium separation, and $\Gamma_0$ is the interfacial adhesion energy per unit area. As plotted in Fig. 1b, the interaction potential reaches a minimum at $z = h_0$, and the adhesion energy ($\Gamma_0$) corresponds to the depth of the energy well at the equilibrium separation. Therefore, the flat monolayer-substrate interaction is fully characterized by the two parameters, $h_0$ and $\Gamma_0$, which are related to the pairwise Lennard-Jones constants in Eq. (1) as $h_0 = \left(\frac{2C_2}{5C_1}\right)^{1/6}$ and $\Gamma_0 = \frac{\pi \rho_s \rho_g C_1}{9 h_0^3}$.

The equilibrium separation ($h_0$) between a monolayer graphene and an oxide substrate is often assumed to be similar to the interlayer spacing in bulk graphite, i.e., 0.34 nm. However, atomic force microscopy (AFM) measurements of graphene on SiO$_2$ have reported values ranging from 0.4 to 0.9 nm [4, 21, 22], commonly referred to as the thickness of monolayer graphene. On the other hand, the adhesion energy ($\Gamma_0$) for monolayer graphene on oxide has not been experimentally measured to the best of our knowledge, while a value of 0.6 eV/nm$^2$ was estimated based on the interlayer interaction energy in graphite [4]. In the present study, we take $h_0 = 0.6$ nm and $\Gamma_0 = 0.6$ eV/nm$^2$ (or equivalently, 0.096 J/m$^2$) as the representative values in all calculations.

The van der Waals interaction force (per unit area, positive for attraction) between a flat graphene monolayer and a flat substrate surface can be obtained by taking the first derivative of the interaction energy in Eq. (3) with respect to $z$, and the corresponding stiffness is obtained by taking the second derivative, namely

$$\sigma_{vdW}(z) = \frac{dU_{vdW}}{dz} = \frac{9\Gamma_0}{2h_0}\left[\left(\frac{h_0}{z}\right)^4 - \left(\frac{h_0}{z}\right)^{10}\right], \tag{4}$$

$$k_{vdW}(z) = \frac{d^2 U_{vdW}}{dz^2} = \frac{27\Gamma_0}{h_0^2}\left[-\frac{2}{3}\left(\frac{h_0}{z}\right)^5 + \frac{5}{3}\left(\frac{h_0}{z}\right)^{11}\right]. \tag{5}$$

As plotted in Fig. 1c, the van der Waals force ($\sigma_{vdW}$) is zero at the equilibrium separation ($z = h_0$) and reaches a maximum (thus termed as *strength*), $\sigma_{max} = 1.466 \frac{\Gamma_0}{h_0}$, at $z = 1.165 h_0$. The initial stiffness at the equilibrium separation is: $k_0 = \frac{27 \Gamma_0}{h_0^2}$. Taking $h_0 = 0.6$ nm and $\Gamma_0 = 0.6$ eV/nm², we obtain that $\sigma_{max} = 230$ MPa and $k_0 = 7200$ MPa/nm. Alternatively, the values for $h_0$ and $\Gamma_0$ may be estimated from mechanical measurements of the initial stiffness and the strength of graphene-substrate interaction (i.e., $h_0 \sim \sigma_{max}/k_0$ and $\Gamma_0 \sim \sigma_{max}^2/k_0$).

B. *Strain-induced instability*

Subject to an in-plane compressive strain, the flat graphene monolayer may become unstable and develop corrugations. The compressive strain may result from thermal expansion mismatch between graphene and the substrate, noting in particular that the graphene has a negative thermal expansion coefficient over a large temperature range [14]. Corrugation reduces the elastic strain energy in the graphene monolayer, which consists of two parts, one for in-plane compression and the other for bending; the former decreases and the latter increases upon corrugation. In addition, corrugation increases the van der Waals interaction energy between graphene and the substrate. The competition among the energetic terms sets a critical strain for onset of strain-induced corrugation as well as the equilibrium corrugation amplitude and wavelength beyond the critical strain.

Assume a sinusoidal corrugation of the graphene monolayer in form of

$$z_g(x) = h_0 + \delta_g \sin\left(\frac{2\pi x}{\lambda}\right), \qquad (6)$$

where $\delta_g$ is the corrugation amplitude and $\lambda$ the wavelength (Fig. 2a). To the leading orders of the corrugation amplitude, the elastic strain energy per unit area of graphene is

$$\tilde{U}_g \approx \left[\frac{C\varepsilon}{4}\left(\frac{2\pi}{\lambda}\right)^2 + \frac{D}{4}\left(\frac{2\pi}{\lambda}\right)^4\right]\delta_g^2 + \frac{3C}{64}\left(\frac{2\pi}{\lambda}\right)^4\delta_g^4, \tag{7}$$

where $\varepsilon$ is the mismatch strain ($\varepsilon < 0$ for a compressive strain, relative to the ground state of graphene), $C$ is the 2-D in-plane elastic modulus, and $D$ is the bending modulus of graphene. The derivation of Eq. (7) is similar to that for thin film wrinkling [23], except for the fact that the bending modulus of monolayer graphene is not directly related to the in-plane modulus by the classical elastic plate theory [13]. Instead, both $C$ and $D$ are intrinsic elastic properties of the graphene lattice. Based on previous first-principle calculations [24], we take $C$ = 353 N/m and $D$ = 0.238 nN-nm (~1.5 eV) for the graphene monolayer. Under the condition of small deformation, the elastic properties of graphene are linear and isotropic [25].

The van der Waals interaction energy between the corrugated monolayer and the substrate can be obtained by integrating the atom-surface interaction energy over one wavelength of the sinusoidal corrugation. For a relatively long corrugation wavelength, compared to the C-C- bond length (~0.142 nm) in graphene, the average interaction energy per unit area is approximately

$$\tilde{U}_{vdW} = \frac{1}{\lambda}\int_0^\lambda U_{vdW}(z_g)dx \approx \Gamma_0\left(-1 + \frac{27}{4}\frac{\delta_g^2}{h_0^2} + \frac{675}{8}\frac{\delta_g^4}{h_0^4}\right). \tag{8}$$

Combining Eqs. (7) and (8), the total free energy (per unit area) to the leading orders of the corrugation amplitude is

$$\tilde{U}_{total} = \Gamma_0\left(-1 + \frac{\delta_g^2}{4h_0^2}\left[27 + \frac{C\varepsilon}{\Gamma_0}\left(\frac{2\pi h_0}{\lambda}\right)^2 + \frac{D}{\Gamma_0 h_0^2}\left(\frac{2\pi h_0}{\lambda}\right)^4\right] + \frac{\delta_g^4}{8h_0^4}\left[675 + \frac{3C}{8\Gamma_0}\left(\frac{2\pi h_0}{\lambda}\right)^4\right]\right). \quad (9)$$

The equilibrium wavelength and amplitude for the strain-induced corrugation are then determined by minimizing the total energy. First, by setting $\frac{\partial \tilde{U}_{total}}{\partial \delta_g} = 0$, we obtain that

$$\left(\frac{\delta_g}{h_0}\right)^2 = \frac{-27 - \frac{C\varepsilon}{\Gamma_0}\left(\frac{2\pi h_0}{\lambda}\right)^2 - \frac{D}{\Gamma_0 h_0^2}\left(\frac{2\pi h_0}{\lambda}\right)^4}{675 + \frac{3C}{8\Gamma_0}\left(\frac{2\pi h_0}{\lambda}\right)^4}. \quad (10)$$

By setting $\frac{\partial \tilde{U}_{total}}{\partial \lambda} = 0$, we obtain that

$$\varepsilon\left(\frac{\lambda}{2\pi h_0}\right)^2 + \frac{2D}{Ch_0^2} + \frac{3\delta_g^2}{8h_0^2} = 0. \quad (11)$$

Inserting Eq. (10) into Eq. (11), we obtain a nonlinear equation for the corrugation wavelength:

$$\frac{400D}{Ch_0^2}\frac{\varepsilon}{\varepsilon_c}\left(\frac{\lambda}{\lambda_c}\right)^6 + \left(3 - \frac{400D}{Ch_0^2}\right)\left(\frac{\lambda}{\lambda_c}\right)^4 - 3 = 0, \quad (12)$$

where

$$\lambda_c = 2\pi\left(\frac{Dh_0^2}{27\Gamma_0}\right)^{1/4}, \quad (13)$$

$$\varepsilon_c = -\frac{6\sqrt{3\Gamma_0 D}}{Ch_0}. \quad (14)$$

Solving Eq. (12) gives the corrugation wavelength of graphene as a function of the mismatch strain (Fig. 2b), while the corresponding corrugation amplitude (Fig. 2c) is calculated from Eq. (10). We note that only when $\frac{\varepsilon}{\varepsilon_c} \geq 1$ does there exist a real-valued solution for the corrugation amplitude. Thus, $\varepsilon_c$ is the critical mismatch strain, beyond which the graphene monolayer becomes corrugated even on a perfectly flat substrate surface. Using the representative values ($C$ = 353 N/m, $D$ = 0.238 nN-nm, $h_0$ = 0.6 nm, and $\Gamma_0$ = 0.6 eV/nm$^2$), we obtain that $\varepsilon_c = -0.0074$ and $\lambda_c = 2.68$ nm. Therefore, a small amount of compressive mismatch strain (< 1%) is sufficient to make the monolayer corrugate due to the strain-induced instability. It can be seen in Fig. 2 that $\lambda = \lambda_c$ and $\delta_g = 0$ when $\frac{\varepsilon}{\varepsilon_c} = 1$. When $\frac{\varepsilon}{\varepsilon_c} > 1$, the corrugation wavelength decreases and the amplitude increases with the magnitude of the compressive strain, similar to wrinkling of an elastic thin film on a hyperelastic substrate [26]. The predicted wavelength (~ 2 nm) for the strain-induced corrugation is much longer than the interatomic bond length of graphene (~0.142 nm), thus justifying the treatment of the graphene monolayer as a continuum membrane in the present study. Noticeably, the strain-induced corrugation wavelength is several times shorter than the reported wavelength (5-10 nm) for the intrinsic ripples in suspended graphene sheets [1, 2], while the corrugation amplitude is comparable to the height fluctuation (~0.07 nm) in atomistic Monte Carlo simulations [2].

III. **Graphene on a corrugated surface**

In this section, we consider effects of substrate surface corrugation on morphology of supported graphene. First we derive an approximate interaction energy function between a single atom and a substrate with a periodically corrugated surface. The interaction energy between a graphene monolayer and a corrugated surface is then obtained analytically to the leading order of the corrugation amplitudes, based on which the equilibrium morphology and the effects of mismatch strain are analyzed.

A. *Effect of surface corrugation on atom-substrate interaction*

Consider the van der Waals interaction between a single carbon atom and a substrate with a corrugated surface (Fig. 3a). Assume the surface to be sinusoidal with an amplitude ($\delta_s$) and a wavelength ($\lambda$), namely

$$z_s(x) = \delta_s \sin \frac{2\pi x}{\lambda}. \tag{15}$$

Let $(x_0, y_0, z_0)$ denote the position of a carbon atom near the corrugated surface. The van der Waals interaction energy is obtained by integrating the pairwise Lennard-Jones potential in Eq. (1) over the volume of the substrate (see Appendix for details). To the leading orders of the surface corrugation amplitude, we obtain that

$$W(x_0, z_0) \approx W_0(z_0) + W_1(z_0)\sin\left(\frac{2\pi x_0}{\lambda}\right)\left(\frac{\delta_s}{z_0}\right) + \left[W_2(z_0) - W_3(z_0)\cos\left(\frac{4\pi x_0}{\lambda}\right)\right]\left(\frac{\delta_s}{z_0}\right)^2, \tag{16}$$

where

$$W_0(z_0) = -\frac{\Gamma_0}{\rho_g}\left[\frac{3}{2}\left(\frac{h_0}{z_0}\right)^3 - \frac{1}{2}\left(\frac{h_0}{z_0}\right)^9\right], \tag{17}$$

$$W_1(z_0) = -\frac{9\pi^2\Gamma_0}{\rho_g}\left[\frac{h_0^3}{z_0\lambda^2}K_2\left(\frac{2\pi z_0}{\lambda}\right) - \frac{\pi^3 h_0^9}{24 z_0^4 \lambda^5}K_5\left(\frac{2\pi z_0}{\lambda}\right)\right], \tag{18}$$

$$W_2(z_0) = -\frac{9\Gamma_0}{2\rho_g}\left[\left(\frac{h_0}{z_0}\right)^3 - \frac{5}{2}\left(\frac{h_0}{z_0}\right)^9\right], \tag{19}$$

$$W_3(z_0) = -\frac{36\pi^3\Gamma_0}{\rho_g}\left[\frac{h_0^3}{\lambda^3}K_3\left(\frac{4\pi z_0}{\lambda}\right) - \frac{\pi^3 h_0^9}{3 z_0^3 \lambda^6}K_6\left(\frac{4\pi z_0}{\lambda}\right)\right], \tag{20}$$

and $K_n(z)$ is the modified Bessel function of the second kind.

Evidently, the first term on the right hand side of Eq. (16) is the interaction energy between a single atom and a substrate with a perfectly flat surface, with an equilibrium separation $h_0$ and the binding energy $\frac{\Gamma_0}{\rho_g}$ (per atom), which naturally leads to the flat monolayer-substrate interaction energy (per unit area) in Eq. (3). The second term oscillates with the in-plane coordinate $x_0$, linearly proportional to the surface corrugation, while the third term represents a second-order effect on the interaction energy.

Derivatives of the interaction energy with respect to the coordinates of the carbon atom give the van der Waals forces acting on the atom in $z$ and $x$ directions, respectively:

$$F_z(x_0, z_0) \approx \frac{dW_0}{dz_0} + \left(\frac{dW_1}{dz_0} - \frac{W_1}{z_0}\right)\sin\left(\frac{2\pi x_0}{\lambda}\right)\left(\frac{\delta_s}{z_0}\right)$$
$$+ \left[\frac{dW_2}{dz_0} - \frac{2W_2}{z_0} - \left(\frac{dW_3}{dz_0} - \frac{2W_3}{z_0}\right)\cos\left(\frac{4\pi x_0}{\lambda}\right)\right]\left(\frac{\delta_s}{z_0}\right)^2, \tag{21}$$

$$F_x(x_0, z_0) \approx -\frac{2\pi W_1}{\lambda}\cos\left(\frac{2\pi x_0}{\lambda}\right)\left(\frac{\delta_s}{z_0}\right) - \frac{4\pi W_3}{\lambda}\sin\left(\frac{4\pi x_0}{\lambda}\right)\left(\frac{\delta_s}{z_0}\right)^2. \tag{22}$$

The equilibrium position of a single atom near the corrugated surface can then be found by setting both the forces to be zero, or equivalently, by minimizing the interaction energy in Eq. (16).

By retaining only the first-order terms in Eq. (21), we obtain the equilibrium $z$-coordinate as a function of the $x$-coordinate of the carbon atom, namely

$$z_0^*(x_0) \approx h_0 \left[1 + \beta \left(\frac{\delta_s}{h_0}\right) \sin \frac{2\pi x_0}{\lambda}\right], \quad (23)$$

where

$$\beta = \frac{2\pi^3}{3} \left[-\frac{h_0^3}{\lambda^3} K_3 \left(\frac{2\pi h_0}{\lambda}\right) + \frac{\pi^3 h_0^6}{24\lambda^6} K_6 \left(\frac{2\pi h_0}{\lambda}\right)\right]. \quad (24)$$

Furthermore, the first term on the right hand side of Eq. (22) suggests that the in-plane force ($F_x$) vanishes when $\frac{2\pi x_0}{\lambda} = 2n\pi \pm \frac{\pi}{2}$ ($n$ is an integer), i.e., when the atom is directly above a peak or a trough of the surface corrugation. However, examination of the stability shows that only the equilibrium positions above the trough of the corrugation is stable, because $W_1(h_0) > 0$ and the atom-substrate interaction energy reaches a minimum when $\frac{2\pi x_0}{\lambda} = 2n\pi - \frac{\pi}{2}$. Including the second-order terms does not affect the equilibrium $x$-coordinate, but does affect the equilibrium $z$-coordinate as well as the energy magnitude (binding energy) due to the nonlinear nature of the van der Waals interaction. As shown in the next section, the second-order terms are essential in considering monolayer-substrate interaction with a corrugated surface.

To illustrate the first-order effect of the substrate surface corrugation on the atom-substrate interaction, Figure 3b plots the dimensionless parameter $\beta$ as a function of the corrugation wavelength. As given in Eq. (23), the equilibrium separation of a single atom from the corrugated surface oscillates with the $x$-coordinate with an amplitude $\beta \delta_s$, where $\beta$ depends only on the ratio, $\lambda/h_0$. The value of $\beta$ has two limits: for a long-wavelength surface corrugation ($\lambda \gg h_0$), $\beta \approx 1$, meaning that the equilibrium position ($z_0^*$) by Eq. (23) follows exactly the surface corrugation with a constant separation $h_0$; on the other hand, for a short-wavelength corrugation ($\frac{\lambda}{h_0} < 0.5$), $\beta \approx 0$, meaning that the equilibrium position ($z_0^*$) is independent of the surface corrugation (i.e., the surface is effectively flat). For surface corrugation of intermediate wavelengths, $\beta$ varies between 0 and 1, and the oscillation of the equilibrium position ($z_0^*$) has a smaller amplitude ($\beta \delta_s$) than the substrate surface corrugation. Thus, it is expected that a monolayer of atoms (e.g., graphene) on a corrugated surface would in general be corrugated, partly conforming to the surface with a smaller amplitude, except for an effectively flat surface with very short corrugation wavelengths.

B. *Surface induced corrugation of graphene*

When a graphene monolayer is placed on top of a corrugated substrate surface, the van der Waals interaction between the individual carbon atoms and the substrate favors conformal corrugation of the monolayer. However, the corrugation-induced deformation of the monolayer increases the elastic strain energy in the monolayer, which tends to resist corrugation. The equilibrium morphology of the supported graphene monolayer is thus determined by the competition between the van der Waals interaction and the intrinsic elasticity of graphene, assuming the substrate to be rigid. As illustrated in Fig. 4, on a periodically corrugated surface as described by Eq. (15), we assume a sinusoidal morphology for the monolayer:

$$z_g(x) = h + \delta_g \sin\left(\frac{2\pi x}{\lambda}\right), \tag{25}$$

where $h$ is the mean separation (or thickness) between the substrate and monolayer, and $\delta_g$ is the corrugation amplitude of the monolayer. Unlike Eq. (6), here the mean separation ($h$) does not necessarily equal $h_0$. Instead, both $h$ and $\delta_g$ depend on the wavelength ($\lambda$) and amplitude ($\delta_s$) of the surface corrugation, which are to be determined by minimizing the total free energy of the system, including contributions from both the van der Waals interaction and the elastic deformation of the supported graphene.

As given in Eq. (7), the elastic strain energy per unit area of graphene ($\widetilde{U}_g$) consists of bending and in-plane components, which is independent of the mean separation ($h$). The van der Waals interaction energy between the corrugated monolayer and the substrate is obtained by integrating Eq. (16) over one wavelength of the periodic corrugation. To the leading order of the corrugation amplitudes, the monolayer-surface interaction energy (per unit area) is

$$\widetilde{U}_{vdW}(h, \delta_g) = \frac{\rho_g}{\lambda} \int_0^\lambda W\left(x, z_g(x)\right) dx$$

$$\approx U_{vdW}(h) + U_1(h)\left[\left(\frac{\delta_g}{h_0}\right)^2 + \left(\frac{\delta_s}{h_0}\right)^2\right] + U_2(h)\frac{\delta_g \delta_s}{h_0^2}, \tag{26}$$

where

$$U_1(h) = \frac{9\Gamma_0}{2}\left[-\left(\frac{h_0}{h}\right)^5 + \frac{5}{2}\left(\frac{h_0}{h}\right)^{11}\right] = \frac{h_0^2}{4} k_{vdW}(h), \tag{27}$$

$$U_2(h) = 9\pi^3 \Gamma_0 \left[\frac{h_0^5}{\lambda^3 h^2} K_3\left(\frac{2\pi h}{\lambda}\right) - \frac{\pi^3 h_0^{11}}{24 h^5 \lambda^6} K_6\left(\frac{2\pi h}{\lambda}\right)\right]. \tag{28}$$

We note that the leading order perturbation to the monolayer-surface interaction energy is quadratic with respect to the corrugation amplitudes ($\delta_s$ and $\delta_g$), including a coupling term ($\sim \delta_s \delta_g$). The second-order effect in Eq. (16) is thus essential for the consideration of monolayer-substrate interaction. The total free energy of the corrugated monolayer-substrate system is then,

$$\widetilde{U}_{total}(h, \delta_g) = \widetilde{U}_{vdW}(h, \delta_g) + \widetilde{U}_g(\delta_g). \tag{29}$$

First consider two limiting cases, respectively, by assuming the monolayer to be perfectly flat (i.e., $\delta_g = 0$) and to be fully conformal to the substrate surface (i.e., $\delta_g = \delta_s$). For each case, the total free energy is a function of $h$, which can be easily minimized. Given an amplitude of the substrate surface corrugation ($\delta_s$), the free energy for a flat monolayer is independent of the wavelength ($\lambda$), while the free energy for the conformal monolayer increases as the wavelength decreases. A comparison of the minimum energy between these two cases (Fig. 5) shows that the flat monolayer is energetically more favorable at small wavelengths while the conformal monolayer is favorable at long wavelengths. The critical wavelength for the cross-over depends on the surface corrugation amplitude ($\delta_s$). This comparison suggests a transition from fully conformal to flat morphology for the monolayer as the wavelength decreases. In general, the monolayer is likely to be partly conformal to the surface corrugation with a corrugation amplitude in between of the two limits (i.e., $0 < \delta_g < \delta_s$). Here we have set the mismatch strain ($\varepsilon$) to be zero, postponing discussions of its effect till the next section.

We now determine the equilibrium mean thickness ($h$) and the corrugation amplitude ($\delta_g$) simultaneously for the graphene monolayer, by minimizing the total free energy in Eq. (29)

with specific parameters for the substrate surface corrugation ($\delta_s$ and $\lambda$). First, by setting $\frac{\partial \widetilde{U}_{total}}{\partial \delta_g} = 0$, we obtain that

$$\frac{\delta_g}{\delta_s} = \frac{-U_2(h)}{2U_1(h) + \frac{D}{2h_0^2}\left(\frac{2\pi h_0}{\lambda}\right)^4 + \frac{C\varepsilon}{2}\left(\frac{2\pi h_0}{\lambda}\right)^2} = f\left(\frac{h}{h_0}; \frac{\lambda}{h_0}, \frac{D}{\Gamma_0 h_0^2}, \frac{C\varepsilon}{\Gamma_0}\right). \tag{30}$$

By setting $\frac{\partial \widetilde{U}_{total}}{\partial h} = 0$, we obtain that

$$\sigma_{vdW}(h) + U_1'(h)\left[\left(\frac{\delta_g}{h_0}\right)^2 + \left(\frac{\delta_s}{h_0}\right)^2\right] + U_2'(h)\frac{\delta_g \delta_s}{h_0^2} = 0, \tag{31}$$

where a prime indicates derivative with respect to *h*. By substituting Eqs. (30) into Eq. (31), we obtain a nonlinear equation for the mean thickness *h*, which is then solved numerically. The result can be written in a dimensionless form, namely

$$\frac{h}{h_0} = g\left(\frac{\lambda}{h_0}, \frac{\delta_s}{h_0}; \frac{D}{\Gamma_0 h_0^2}, \frac{C\varepsilon}{\Gamma_0}\right), \tag{32}$$

which is plotted in Fig. 6a. The corresponding corrugation amplitude of the monolayer is then calculated by inserting Eq. (32) into Eq. (30), as plotted in Fig. 6b. Here the dimensionless parameter $\frac{D}{\Gamma_0 h_0^2} = 6.94$ and the mismatch strain $\varepsilon = 0$. Limited by the leading-order approximations in the free energy function, relatively small corrugation amplitudes (e.g., $\frac{\delta_s}{h_0} < 0.5$) are considered.

Figure 6 shows clearly the transition from a long-wavelength limit to a short-wavelength limit for both the mean separation (*h*) and the surface-induced corrugation amplitude ($\delta_g$) of the supported graphene monolayer. At the limit of long wavelengths (e.g., $\frac{\lambda}{h_0} > 15$), the mean thickness approaches $h_0$ and the corrugation amplitude of graphene approaches that of the substrate surface ($\frac{\delta_g}{\delta_s} \to 1$). Thus, the monolayer morphology is completely conformal to the surface. However, as the corrugation wavelength decreases, the mean separation increases and approaches a plateau at the short-wavelength limit that depends on the surface corrugation amplitude. Meanwhile, the corrugation amplitude of graphene decreases and approaches zero ($\frac{\delta_g}{\delta_s} \to 0$). Therefore, the graphene tends to be partly conformal to the surface corrugation for intermediate wavelengths and becomes completely non-conformal (flat) for short wavelengths. The transition process is increasingly abrupt as the surface corrugation amplitude increases, and a snap-through instability [19] is predicted for relatively large amplitudes (not shown in Fig. 6). As a quantitative example, for a surface amplitude $\delta_s = 0.24$nm ($\frac{\delta_s}{h_0} = 0.4$), the transition wavelength is about 4 nm ($\frac{\lambda}{h_0} \sim 7$) and the supported graphene monolayer becomes almost completely conformal to the surface corrugation for wavelengths longer than 9 nm ($\frac{\lambda}{h_0} > 15$).

Corrugation of monolayer graphene supported on SiO$_2$ has been probed experimentally [4-6]. Compared to the bare oxide surface, the supported graphene was found to have a smaller root-mean-square roughness and a longer correlation length [4], both of which can be qualitatively understood based on the theoretical results in the present study. As shown in Fig. 6b, the amplitude ratio ($\frac{\delta_g}{\delta_s}$) is less than 1 in general, suggesting a flatter morphology for the supported graphene. One may consider a randomly corrugated substrate surface consisting of many Fourier components with statistically distributed wavelengths and amplitudes, on top of which the graphene corrugates partly conformal to the long-wavelength Fourier modes; The short-wavelength modes of surface corrugation are filtered out. Consequently, the average corrugation wavelength in graphene is longer than that for the bare substrate surface, and the root-mean-square amplitude is smaller. Similar results were reported in a previous study [19], where the van der Waals interaction energy was calculated numerically based on a Monte Carlo method. There, the corrugation amplitude of the substrate surface was set to a fixed value (0.5 nm), and the amplitude ratio and the equilibrium separation were determined as functions of the ratio, $\frac{\lambda}{\delta_s}$, for different values of the adhesion energy. By the dimensional consideration in Eqs. (30) and (32), we emphasize that the morphology of graphene depends on both the corrugation amplitude and the wavelength of the substrate surface, as shown clearly in Fig. 6.

In addition to the effects on the morphology, we find that the effective adhesion energy between a graphene monolayer and an oxide substrate also depends on the surface corrugation. The magnitude of the total free energy ($\widetilde{\widetilde{U}}_{total}$) corresponding to the equilibrium mean separation and the equilibrium corrugation amplitude gives a measure of the effective adhesion energy, which includes the contributions from both the van der Waals interaction and the elastic strain energy in the corrugated graphene monolayer. Figure 7 shows the normalized adhesion energy as a function of the corrugation wavelength for different amplitudes of the substrate surface corrugation. The adhesion energy approaches $\Gamma_0$ at the long-wavelength limit, but decreases as the wavelength decreases, approaching a plateau at the short-wavelength limit. The adhesion energy decreases with increasing amplitude of the substrate surface corrugation. For $\frac{\delta_s}{h_0} = 0.4$, the adhesion energy drops nearly 20% from long-wavelength to short-wavelength corrugations. Since the graphene monolayer is predicted to be nearly flat at the short-wavelength limit, the substrate surface may be considered effectively flat, but the larger equilibrium separation ($h > h_0$) leads to a lower adhesion energy. Interestingly, a previous study based on a perturbation theory and molecular dynamics simulations predicted a similar dependence of friction between a substrate and an absorbed monolayer on the surface corrugation [27], suggesting an intimate relationship between adhesion and friction.

C. *Effect of mismatch strain*

As shown in a previous section (II.B), a compressive mismatch strain can cause a supported graphene monolayer to corrugate even on a perfectly flat substrate surface as a result of strain-induced instability. On a corrugated substrate surface, the morphology of the supported graphene depends on both the surface corrugation and the mismatch strain. By minimizing the total free energy in Eq. (29) with a tensile ($\varepsilon > 0$) or compressive ($\varepsilon < 0$) mismatch strain, the effect of mismatch strain is analyzed. As predicted by Eqs. (30) and (32), the mean separation and the corrugation amplitude of graphene depends on the mismatch strain through the dimensionless group, $\frac{C\varepsilon}{\Gamma_0}$. In general, a tensile strain tends to flatten the supported graphene and a compressive strain tends to increase the corrugation amplitude. As shown in Fig. 8a, for a given corrugation wavelength (e.g., $\frac{\lambda}{h_0} = 10$), the amplitude ratio decreases with the increasing tensile strain. For the case of $\frac{\delta_s}{h_0} = 0.4$, a snap-through occurs at $\varepsilon \sim 0.004$, beyond which the graphene becomes flat ($\delta_g = 0$). Under a compressive strain, the amplitude ratio increases until a critical compressive strain, beyond which the strain-induced instability dominates and a higher-order analysis is required to predict the corrugation amplitude for the graphene. The critical compressive strain is determined as a function of the surface corrugation wavelength ($\lambda$) for different corrugation amplitudes ($\delta_s$) in Fig. 9. On a flat substrate surface ($\delta_s = 0$), the critical strain is obtained from Eq. (10) by setting the right-hand side to be zero, which has a minimum ($\varepsilon_c$) at the critical wavelength ($\lambda_c$) as given in Eqs. (14) and (13) respectively. On a corrugated substrate surface, the critical compressive strain decreases as the corrugation amplitude increases, and the corrugation wavelength for the minimum critical strain increases. It is thus predicted that the substrate surface corrugation promotes onset of the strain-induced instability at a lower compressive strain and a longer wavelength. In addition, Figure 8b shows that the mismatch strain, either tensile or compressive, tends to increase the mean thickness of the supported graphene monolayer on a corrugated substrate surface, which can be attributed to the strain-induced change in the corrugation amplitude of graphene. In contrast, on a perfectly flat substrate surface ($\delta_s = 0$), the mean thickness is independent of the mismatch strain. Thus, the effect of mismatch strain is intimately coupled with the effect of substrate surface corrugation.

## IV. Concluding Remarks

In the present study, the effects of substrate surface corrugation and mismatch strain on the morphology of a graphene monolayer supported on an oxide substrate are theoretically analyzed. An analytical approach is presented to explicitly relate the van der Waals interaction energy to the substrate surface corrugation, based on which a transition from conformal to non-conformal morphology is predicted for the supported graphene monolayer. In addition, strain-induced instability is predicted such that a compressive mismatch strain can cause a supported graphene monolayer to corrugate even on a perfectly flat substrate surface. By considering both the substrate surface corrugation and the mismatch strain, it is found that, while a tensile mismatch strain reduces the corrugation amplitude of graphene, a corrugated substrate surface promotes strain-induced instability under a compressive strain.

The implications of the present theoretical results may be taken in two-folds. On one hand, it suggests that an ultraflat graphene monolayer may be achieved either on an ultraflat substrate surface [28] or by imposing a tensile mismatch strain. The ultraflat substrate surface could have long-wavelength, small-amplitude corrugations or ultrashort-wavelength corrugations, while a relatively small tensile mismatch strain could be sufficient to flatten the supported graphene. On the other hand, periodically corrugated graphene monolayers can be obtained with tunable wavelength and amplitude by either substrate surface patterning [19] or strain-induced instability. Sophisticated strain engineering approaches may be devised to achieve various morphological textures in partly supported graphene [14] and on mechanically flexible substrates [29, 30].

**Acknowledgments**

The authors gratefully acknowledge funding of this work by National Science Foundation through Grant No. 0926851.

**Appendix**

We calculate the interaction energy between an atom and a corrugated surface of the substrate (Fig. 3a) as follows:

$$W(x_0, z_0) = \int_{-\infty}^{\infty} \int_{-\infty}^{z_s(x)} \int_{-\infty}^{\infty} W_{LJ}(r)\rho_s dy\, dz\, dx, \tag{33}$$

where $W_{LJ}(r)$ is given in Eq. (1) with $r = \sqrt{(x-x_0)^2 + y^2 + (z-z_0)^2}$ and $z_s(x) = \delta_s \sin\frac{2\pi x}{\lambda}$.

First, integrating with respect to $y$, we obtain that

$$W(x_0, z_0) = \int_{-\infty}^{\infty} \int_{-\infty}^{z_s(x)} \rho_s \left(-\frac{3\pi C_1}{8a^5} + \frac{63\pi C_2}{256 a^{11}}\right) dz\, dx, \tag{34}$$

where $a = \sqrt{(x-x_0)^2 + (z-z_0)^2}$.

Next, integrating with respect to $z$ leads to

$$W(x_0, z_0) = \int_{-\infty}^{\infty} \rho_s \left[ \begin{array}{l} -\frac{3\pi C_1}{8}\left(\frac{z_1}{3b^2(z_1^2+b^2)^{3/2}} + \frac{2z_1}{3b^4(z_1^2+b^2)^{1/2}} + \frac{2}{3b^4}\right) \\ +\frac{63\pi C_2}{256}\left(\begin{array}{l}\frac{z_1}{9b^2(z_1^2+b^2)^{9/2}} + \frac{8z_1}{63b^4(z_1^2+b^2)^{7/2}} + \frac{48z_1}{315b^6(z_1^2+b^2)^{5/2}} \\ +\frac{192z_1}{945b^8(z_1^2+b^2)^{3/2}} + \frac{384z_1}{945b^{10}(z_1^2+b^2)^{1/2}} + \frac{384}{945b^{10}}\end{array}\right) \end{array} \right] dx, \tag{35}$$

where $b^2 = (x-x_0)^2$ and $z_1 = \delta_s \sin\frac{2\pi x}{\lambda} - z_0$.

Finally, assuming $\delta_s \ll z_0$, we integrate Eq. (35) to obtain Eq. (16) as the second-order approximation. To reach the final form of Eqs. (17-20), we have used the integral form of the modified Bessel function of the second kind (Basset function):

$$K_n(z) = \frac{\Gamma\left(n+\frac{1}{2}\right)(2z)^n}{\sqrt{\pi}} \int_0^{\infty} \frac{\cos x}{(z^2+x^2)^{n+1/2}} dx. \tag{36}$$

**List of Figures:**

Figure 1: (a) Schematic illustration of a graphene monolayer on a flat substrate surface; (b) The van der Waals interaction energy per unit area as a function of the separation; (c) The van der Waals force per unit area as a function of the separation.

Figure 2: (a) Schematic illustration for strain-induced corrugation of a graphene monolayer on a flat surface; (b) Equilibrium wavelength and (c) corrugation amplitude of the graphene versus the magnitude of the compressive mismatch strain ($\varepsilon < 0$). The vertical dashed line in (b) indicates the critical strain, $\varepsilon_c = -0.0074$, and the horizontal dashed line indicates the corresponding corrugation wavelength, $\lambda_c = 2.68$ nm.

Figure 3: (a) Schematic illustration of a single atom near a corrugated surface; (b) The first-order effect of surface corrugation on the equilibrium separation of a single atom from the surface.

Figure 4: Schematic illustration of a graphene monolayer on a corrugated substrate surface.

Figure 5: Comparison of the total free energies for the two limiting cases of the graphene morphology (flat vs conformal) on a corrugated surface.

Figure 6: (a) Equilibrium mean thickness of monolayer graphene on a corrugated substrate surface, as a function of the corrugation wavelength for different corrugation amplitudes; (b) The ratio of the corrugation amplitudes between the graphene and the substrate surface.

Figure 7: Effective adhesion Energy for a graphene monolayer on a corrugated substrate surface, as a function of the corrugation wavelength for different corrugation amplitudes.

Figure 8: Effects of in-plane mismatch strain on (a) the corrugation amplitude and (b) the mean thickness of monolayer graphene on a corrugated substrate surface with $\frac{\lambda}{h_0} = 10$. The vertical dashed line indicates the critical strain ($\varepsilon_c = -0.0074$) for strain-induced instability on a perfectly flat substrate surface ($\delta_s = 0$).

Figure 9: Critical compressive strain for strain-induced instability of a supported graphene monolayer.

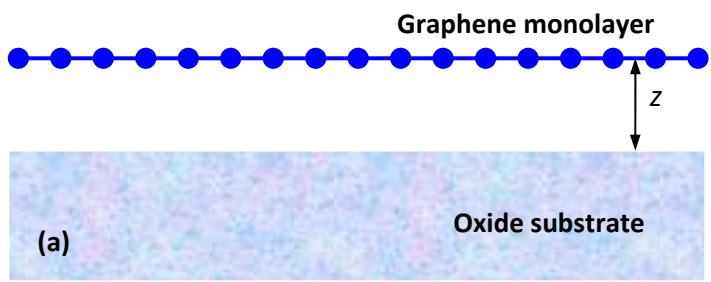

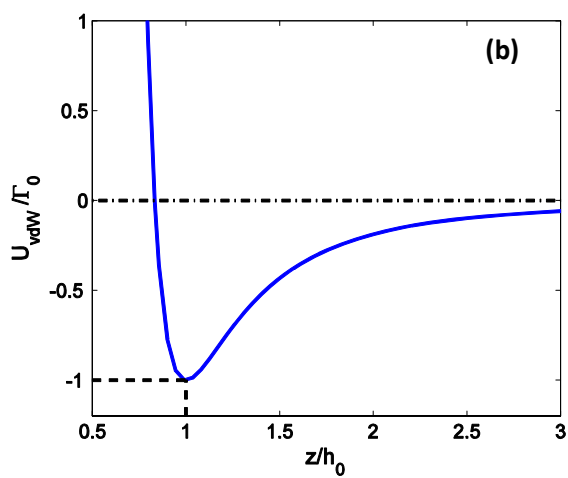

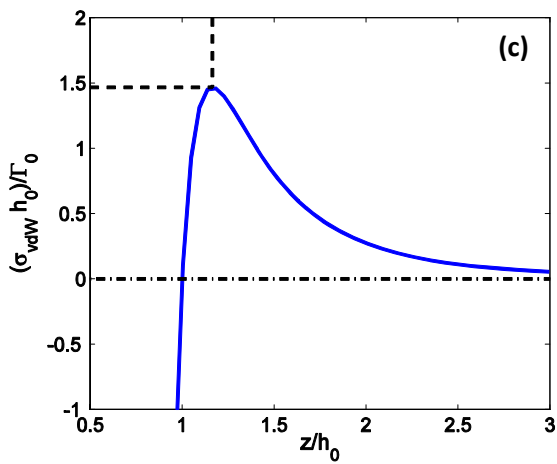

Figure 1: (a) Schematic illustration of a graphene monolayer on a flat substrate surface; (b) The van der Waals interaction energy per unit area as a function of the separation; (c) The van der Waals force per unit area as a function of the separation.

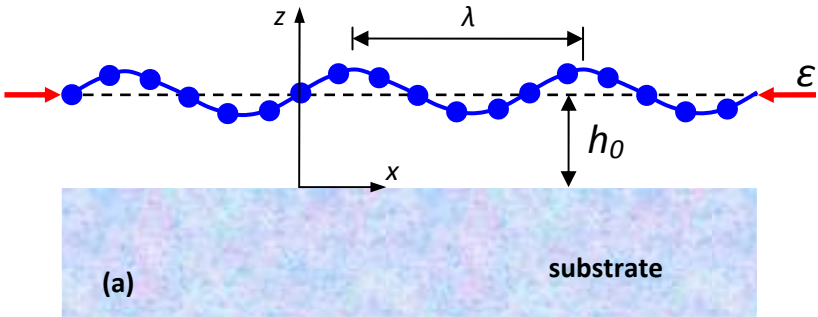

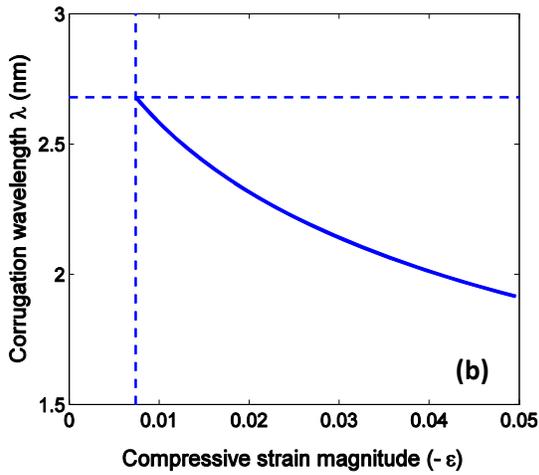

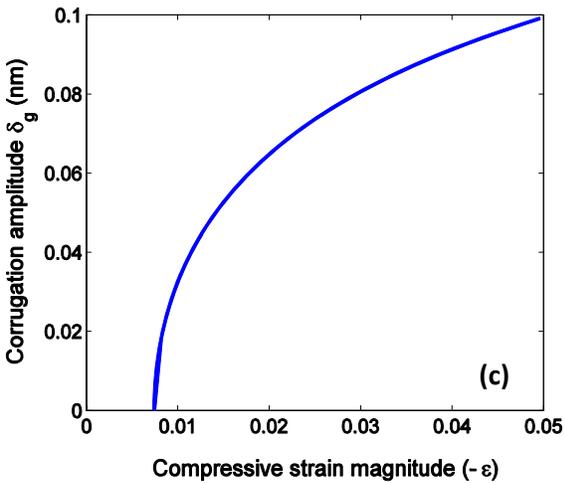

Figure 2: (a) Schematic illustration for strain-induced corrugation of a graphene monolayer on a flat surface; (b) Equilibrium wavelength and (c) corrugation amplitude of the graphene versus the magnitude of the compressive mismatch strain ($\varepsilon < 0$). The vertical dashed line in (b) indicates the critical strain, $\varepsilon_c = -0.0074$, and the horizontal dashed line indicates the corresponding corrugation wavelength, $\lambda_c = 2.68$ nm.

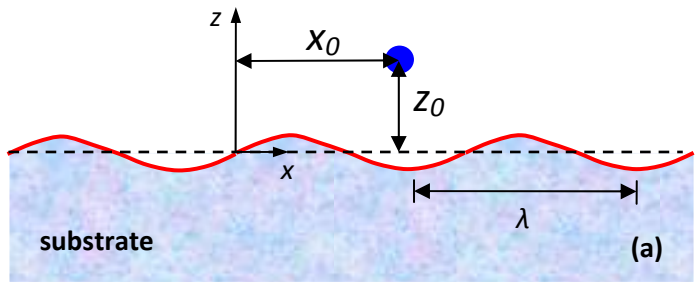

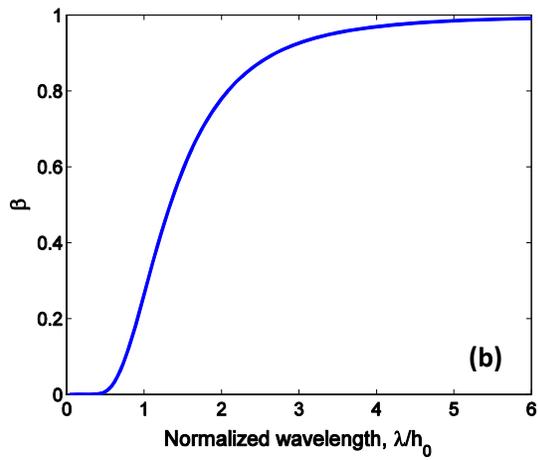

Figure 3: (a) Schematic illustration of a single atom near a corrugated surface; (b) The first-order effect of surface corrugation on the equilibrium separation of a single atom from the surface.

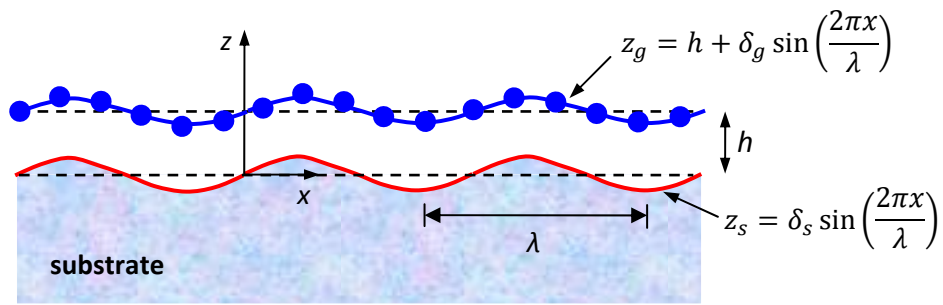

Figure 4: Schematic illustration of a graphene monolayer on a corrugated substrate surface.

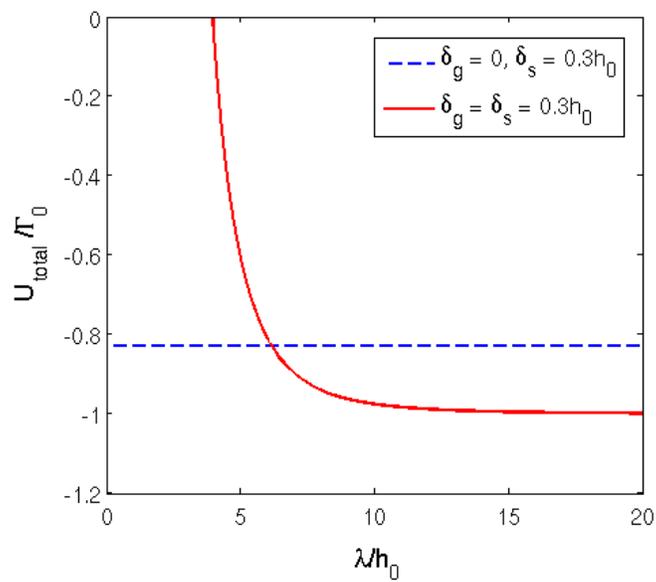

Figure 5: Comparison of the total free energies for the two limiting cases of the graphene morphology (flat vs conformal) on a corrugated surface.

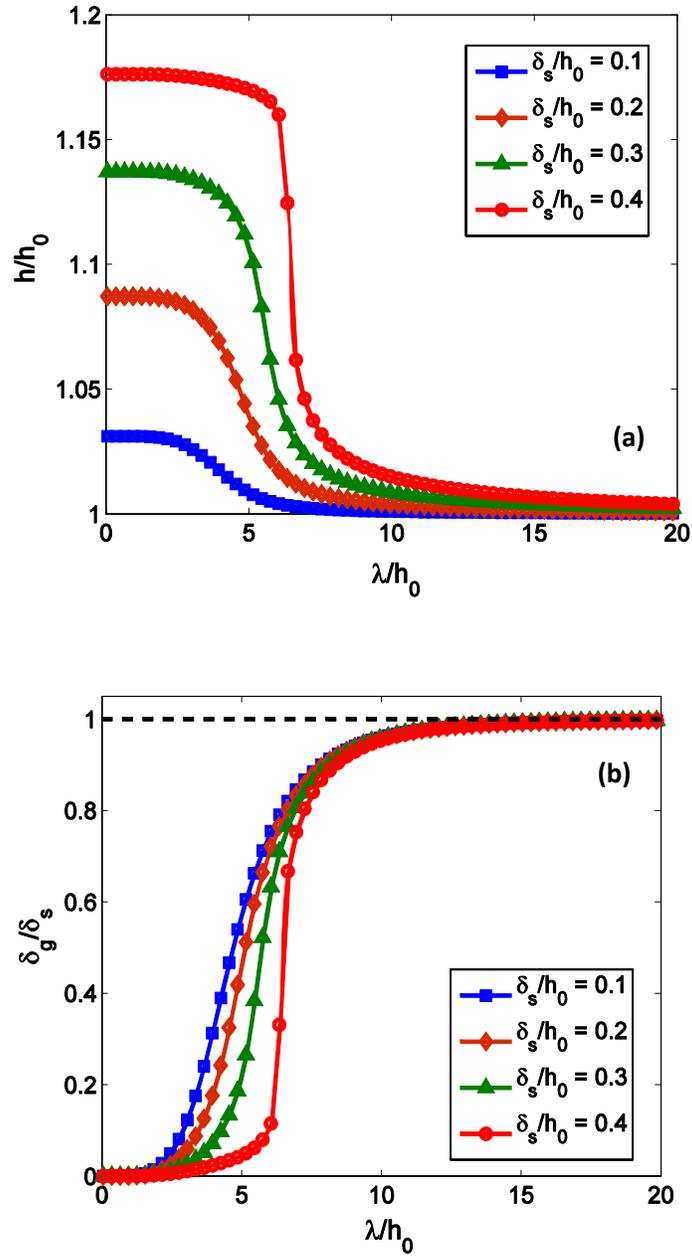

Figure 6: (a) Equilibrium mean thickness of monolayer graphene on a corrugated substrate surface, as a function of the corrugation wavelength for different corrugation amplitudes; (b) The ratio of the corrugation amplitudes between the graphene and the substrate surface.

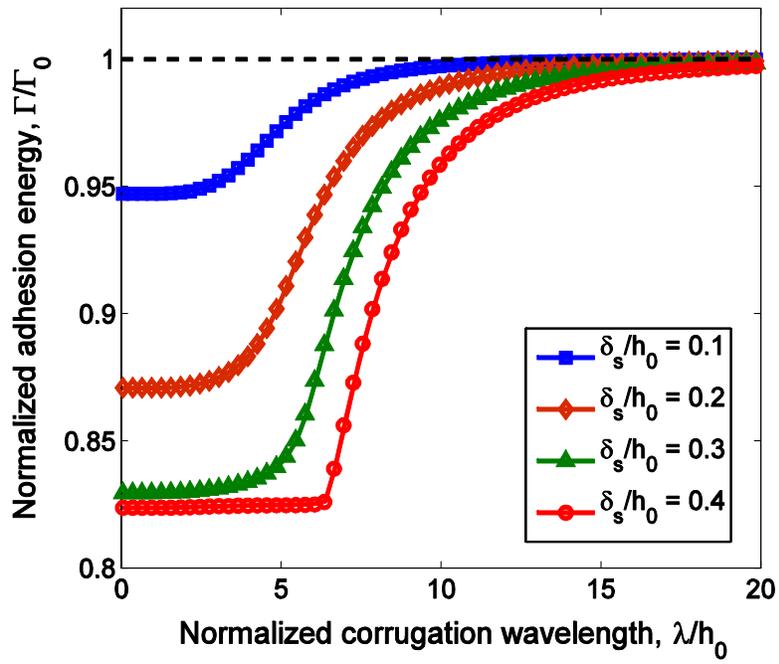

Figure 7: Effective adhesion Energy for a graphene monolayer on a corrugated substrate surface, as a function of the corrugation wavelength for different corrugation amplitudes.

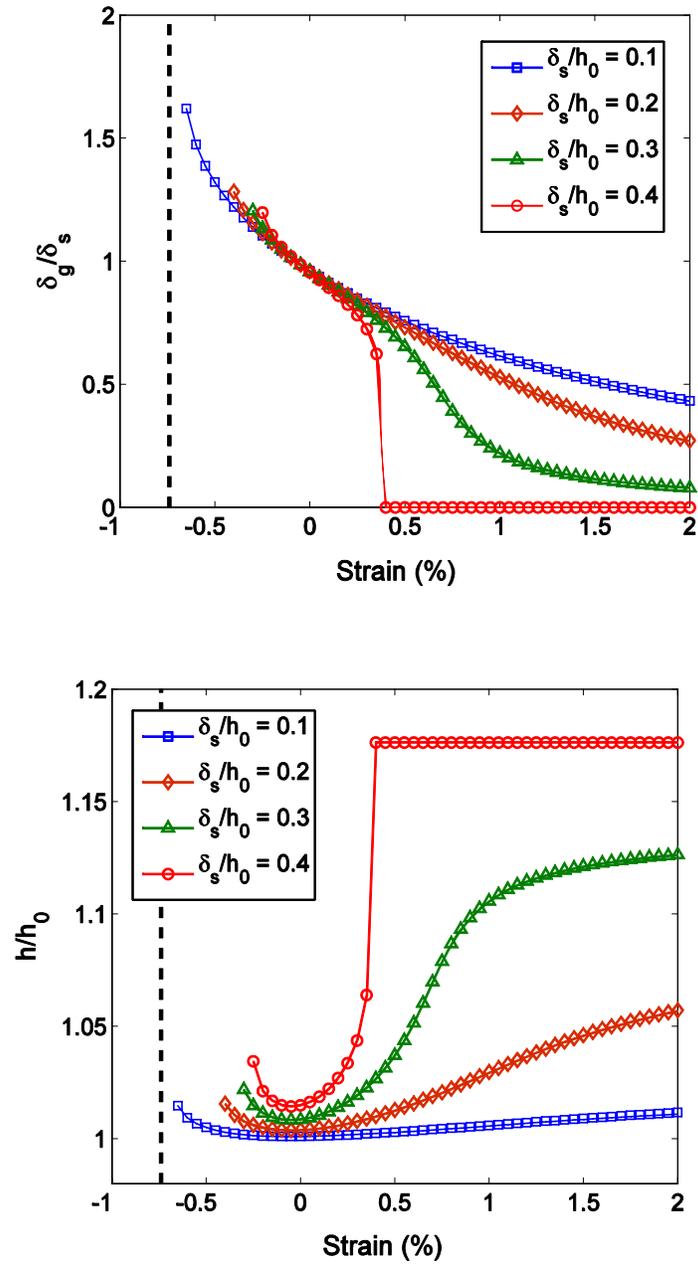

Figure 8: Effects of in-plane mismatch strain on (a) the corrugation amplitude and (b) the mean thickness of monolayer graphene on a corrugated substrate surface with $\frac{\lambda}{h_0} = 10$. The vertical dashed line indicates the critical strain ($\varepsilon_c = -0.0074$) for strain-induced instability on a perfectly flat substrate surface ($\delta_s = 0$).

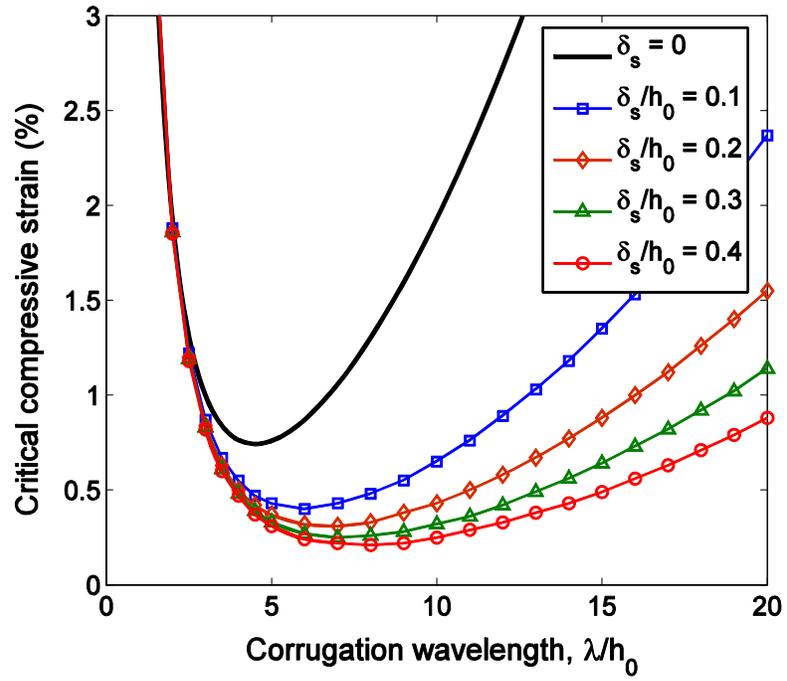

Figure 9: Critical compressive strain for strain-induced instability of a supported graphene monolayer.